\documentclass[twocolumn,aps,pra]{revtex4-2}
\usepackage{bm,bbm}
\usepackage{amsfonts,amsmath,amssymb,latexsym}
\usepackage{graphicx}
\usepackage{mathtools}
\usepackage{xcolor}
\usepackage{hyperref}
\usepackage{physics}
\begin{document} 
\title{Spin-induced and orbital quantum backflow in Pauli theory}
\author{Tomasz Rado\.zycki}
\email{t.radozycki@uksw.edu.pl}
\affiliation{Faculty of Mathematics and Natural Sciences, College of Sciences, Institute of Physical Sciences, Cardinal Stefan Wyszynski University in Warsaw, W\'oycickiego 1/3, 01-938 Warsaw, Poland} 

\begin{abstract}
We show that quantum backflow in spin-$\tfrac{1}{2}$ systems within the Pauli framework can arise from spin-dependent contributions to the probability current. We construct explicit two-mode states in which the orbital current reduces to a uniform drift, so that no interference between longitudinal momentum components is present in this sector.

In such configurations, backflow is generated entirely by spinor structure, which induces time-dependent modulations of the local current leading to flux reversal, while the longitudinal momentum distribution remains strictly supported on positive momenta.

We derive analytical conditions for backflow in terms of spinor and momentum parameters, allowing direct control of the effect via internal degrees of freedom. We further analyze spatial coarse-graining, showing exponential suppression of interference and a qualitatively different scaling of orbital and spin contributions, with spin-induced backflow remaining robust in regimes where orbital effects are strongly suppressed.
\end{abstract}

\maketitle
\section{Introduction}

Quantum backflow is a nonclassical interference effect in which the probability current can become locally negative even when the momentum distribution is restricted to positive values. Introduced by Bracken and Melloy \cite{brack}, it has become a paradigmatic example of quantum transport beyond classical intuition. A central result is a universal upper bound on the integrated backflow, known as the Bracken–Melloy constant \cite{brack,berry}, which has motivated extensive analytical and numerical studies \cite{penz,yearsley,mel2}.

Beyond its conceptual significance, quantum backflow is closely related to theoretical aspects of time-of-arrival measurements, which can be formulated within different frameworks including complex absorbing potentials and operator-based approaches \cite{al1,al2,al3,kij,muga,toa}. More recently, analogous effects have been observed in classical wave systems, including optics, where negative local flux has been both predicted and experimentally demonstrated \cite{berry,saari,kot1,kot2,kot3,elie,daniel,trgaus,zhang,he2}. These results indicate that backflow is a general wave phenomenon rather than a feature specific to quantum mechanics \cite{ibba}.

In the standard Schrödinger setting, backflow arises from interference between momentum components, since the probability current is fully determined by phase gradients. This mechanism extends to relativistic spin-$\tfrac{1}{2}$ systems governed by the Dirac equation, where negative probability current can occur even under restrictions on the momentum support \cite{mell,asfa,ibba}.

However, in relativistic quantum theory, the probability current admits well-established decompositions into orbital and spin parts, in particular through the Gordon and Belinfante constructions \cite{gordon,beli}. Although representation-dependent, these decompositions provide a convenient framework for distinguishing different contributions to the conserved current. This naturally raises the question of whether backflow is solely an interference effect associated with the orbital sector, or whether spin-dependent terms can provide an additional mechanism for local current reversal within a single quantum state.

In the nonrelativistic Pauli regime, where both orbital and spin terms are present, their respective roles in generating negative flux have not been systematically clarified. In particular, it remains unclear whether spin introduces additional degrees of freedom for controlling backflow beyond those available in scalar quantum mechanics, and whether it can support current reversal independently of interference between distinct momentum components.

In this work we show that quantum backflow can occur without longitudinal momentum interference. We construct explicit two-mode states within the Pauli framework in which the orbital current reduces to a uniform drift, while spin-dependent terms generate oscillatory contributions that produce negative values of the total current. This demonstrates that momentum-space interference is not necessary for backflow and identifies spin as an independent control parameter of the effect. Moreover, spin-induced backflow is found to be more robust under spatial coarse-graining, which strongly suppresses orbital interference effects. These results suggest potential relevance for spin-dependent transport phenomena, where control of spin contributions to quantum currents is of practical importance.

The paper is organized as follows. Section~\ref{pacu} introduces the Pauli current with its decomposition into orbital and spin parts and defines the backflow functional. Section~\ref{twom} presents the two-mode construction and establishes the structure of the local current. Section~\ref{explicit} analyzes explicit configurations exhibiting orbital- and spin-induced backflow. Section~\ref{control} discusses parameters controlling the transition between these regimes. Section~\ref{coarg} investigates the effect of spatial coarse-graining and finite detector resolution.

\section{Pauli current and backflow functional}\label{pacu}

Since the key observation of this work is that spin-dependent terms in the Pauli current can generate backflow independently of orbital interference, we start by briefly recalling the probability current in Pauli theory and its decomposition into orbital and spin-dependent contributions. The structure follows from the nonrelativistic limit of the Dirac current \cite{dirac,bjorken,thaller}, and can be obtained, for example, via a Foldy--Wouthuysen transformation \cite{fw} or, equivalently, through the Gordon decomposition \cite{gordon}.

In units $\hbar=c=1$, the free Dirac equation reads
\begin{equation}
(i\gamma^\mu \partial_\mu - m)\psi = 0,
\end{equation}
with the associated conserved current
\begin{equation}
J^\mu = \bar{\psi}\gamma^\mu \psi,
\qquad
\bar{\psi} = \psi^\dagger \gamma^0.
\end{equation}

This current admits a decomposition of the form
\begin{equation}\label{curtotal}
J^\mu =
\frac{1}{2m}\bar{\psi}\left(i\overleftrightarrow{\partial^\mu}\right)\psi
+ \frac{1}{2m}\partial_\nu\!\left(\bar{\psi}\sigma^{\mu\nu}\psi\right)
,
\end{equation}
where
\begin{equation}
\sigma^{\mu\nu} = \frac{i}{2}[\gamma^\mu,\gamma^\nu],
\end{equation}
which yields a formal separation into orbital and spin terms. A closely related and more general decomposition of the energy–momentum tensor, which explicitly reveals the spin contribution to the current, is given by the Belinfante–Rosenfeld construction \cite{beli,rosen}.

Working in the Dirac representation, the bispinor can be written as 
\begin{equation}\label{rozpsi}
\psi = e^{-imt}
\begin{pmatrix}
\phi \\
\chi
\end{pmatrix},
\end{equation}
where the fast rest-energy oscillation has been factored out. This leads to the system of coupled equations
\begin{subequations}\label{comp}
\begin{align}
i\partial_t \phi &= -i\bm{\sigma}\cdot\bm{\nabla}\chi,\\
(i\partial_t + 2m)\chi &= -i\bm{\sigma}\cdot\bm{\nabla}\phi,
\end{align}
\end{subequations}
which constitute the starting point for the nonrelativistic reduction leading to the Pauli equation.

In the nonrelativistic limit, the lower component is eliminated to leading order in $\tfrac{1}{m}$ \cite{sakurai,messiah}:
\begin{equation}\label{wzchi}
\chi = -i\,\frac{\bm{\sigma}\cdot\bm{\nabla}}{2m}\phi + O\Big(\frac{1}{m^2}\Big),
\end{equation}
while the upper component satisfies
\begin{equation}\label{wzphi}
i\partial_t \phi = -\frac{\Delta}{2m}\phi + O\Big(\frac{1}{m^2}\Big),
\end{equation}
which corresponds to the free Schrödinger equation for a Pauli spinor.

Substituting this expansion into the Dirac current, the spatial part of the Pauli current reads
\begin{equation}\label{compcur}
\bm{J}=\bm{J}_o + \bm{J}_s,
\end{equation}
where
\begin{equation}\label{compcurs}
\bm{J}_o = \frac{1}{m}\Im\!\left(\phi^\dagger\bm{\nabla}\phi\right), \quad \bm{J}_s= \frac{1}{m}\,\bm{\nabla}\times\left(\phi^\dagger\frac{\bm{\sigma}}{2}\phi\right).
\end{equation}
The spin current is not uniquely defined, and different physically motivated definitions may differ by divergence-free terms \cite{shi}. In the present work, we adopt the standard Gordon-based separation, which provides a convenient and widely used framework for isolating spin-dependent effects in the current.

This decomposition does not define independent transport channels, but rather a formal partition of the conserved probability current into contributions with different structural origins. Both terms enter linearly and may be analyzed separately in terms of their respective interference properties. 

In what follows the mass factor will be absorbed into the definitions ($\bm{j}=m\bm{J}$), i.e.,
we set $m=1$, so that all quantities become dimensionless:
\begin{subequations}\label{redec}
\begin{align}
\bm{j}_o &= \Im\!\left(\phi^\dagger\bm{\nabla}\phi\right),\\
\bm{j}_s &= \bm{\nabla}\times\left(\phi^\dagger\frac{\bm{\sigma}}{2}\phi\right)
= \bm{\nabla}\times\bm{S},
\end{align}
\end{subequations}
where
\begin{equation}\label{spindens}
\bm{S} = \phi^\dagger \frac{\bm{\sigma}}{2}\phi
\end{equation}
is the spin density \cite{takabayasi,holland}. We further focus on the longitudinal component, i.e., $j_z$, evaluated along a fixed spatial axis. 

In standard formulations of quantum transport, backflow is defined in terms of the probability flux through a surface or finite detector region, often described using positive-operator-valued measures (POVMs) in time-of-arrival approaches \cite{al1,al2,al3,peres}.

In the present work, we adopt a simplified point-detection description, in which the local probability current at a fixed spatial position is used as an effective operational substitute for the flux measured by a sharply localized detector. This approximation is valid when the spatial resolution is small compared to the characteristic wavelength scales of the state. A more detailed analysis of finite spatial resolution and coarse-graining effects is provided in Section~\ref{coarg}.

The backflow functional is then defined as the time-integrated current at a fixed observation point $\bm{r}_0$:
\begin{equation}\label{fff}
F = \int_{t_1}^{t_2} dt\, j_z(\bm{r}_0,t).
\end{equation}

This quantity provides a local diagnostic of probability transport in the idealized limit of a pointlike detector probing the normal component of the current. Although the analysis is formulated in terms of the current evaluated at a single point, the continuity of $j_z(\bm{r},t)$ implies that any local current reversal extends over a finite spatial neighborhood. Thus, negative flux detected at $\bm{r}_0$ is not an isolated point effect but reflects a spatially resolved feature of the probability flow, consistent with finite-resolution detection schemes.

$F$ depends on the full dynamical evolution of the state and therefore captures the local structure of the flow rather than global conservation properties. In particular, negative values of $F$ may occur for suitably constructed states even if the momentum support is restricted to positive values. Such pointwise current-based definitions are standard in studies of quantum backflow and in operational approaches to time-of-arrival problems \cite{brack,yearsley,muga,toa}.

Using (\ref{redec}), the functional $F$ decomposes as
\begin{equation}
F = F_o + F_s,
\end{equation}
with
\begin{subequations}\label{efy}
\begin{align}
F_o &= \int_{t_1}^{t_2} dt\, \Im\!\left(\phi^\dagger \partial_z \phi\right)\Big|_{\bm{r}_0},\\
F_s &= \int_{t_1}^{t_2} dt\, (\nabla \times \bm{S})_z\Big|_{\bm{r}_0}.
\end{align}
\end{subequations}
The orbital term is determined by phase gradients, while the spin term is governed by spatial variations of the spin density. For single-mode states both contributions are trivial: the orbital current reduces to a uniform drift, and the spin term vanishes due to the absence of spatial structure. Substituting a plane-wave Pauli spinor $\phi(\bm{r},t)=u_r e^{i(\bm{p}\cdot\bm{r}-Et)}$, one immediately finds
\begin{equation}
\bm{j}_o = \bm{p}, \qquad \bm{j}_s = 0,
\end{equation}
so that
\begin{equation}
F = (t_2-t_1)p_z.
\end{equation}
Hence no backflow is present, showing that negative flux cannot arise at the single-mode level and must instead originate from superpositions that generate nontrivial spatial structure in the current.

On the other hand, in the case of a superposition $\phi=\sum_j c_j \phi_j$, the current contains interference terms of the form
\begin{equation}
\phi_i^\dagger \nabla \phi_j, \qquad
\phi_i^\dagger \bm{\sigma}\phi_j,
\end{equation}
which generate oscillatory contributions in both space and time and are responsible for sign changes in the local flux.

In scalar quantum mechanics the (Schr\"odinger) current is purely orbital,
\begin{equation}
\bm{j}_{Sch} = \Im(\psi^*\nabla\psi),
\end{equation}
so backflow is entirely driven by momentum-space interference. The Pauli theory introduces an additional spin-mediated channel that can contribute to flux reversal at a fixed point.
In the following sections, explicit solutions of the free Pauli equation are used to disentangle these mechanisms. The orbital contribution requires phase interference, whereas the spin contribution can generate negative flux through the structure of the spin density alone.

\section{Two-mode interference as the mechanism of backflow}\label{twom}

The simplest realization of quantum backflow in the Pauli theory is given by a superposition of two plane-wave solutions. This setting captures both orbital and spin interference effects and admits a fully analytical treatment.

\subsection{Two-mode Pauli state}

Let us consider a superposition of two exact solutions of the free Pauli equation,
\begin{equation}\label{super}
\phi(\bm{r},t) = \phi_1(\bm{r},t) + \phi_2(\bm{r},t),
\end{equation}
with (recall that $\hbar=c=m=1$)
\begin{equation}\label{phij}
\phi_j = c_j\, u_j\, e^{i(\bm{p}_j\cdot \bm{r} - E_j t)},
\qquad
E_j = \frac{\bm{p}_j^2}{2},
\end{equation}
and normalization
\begin{equation}
|c_1|^2 + |c_2|^2 = 1.
\end{equation}

Plane-wave superpositions are used for analytical convenience; however, the resulting mechanism is local in nature and extends to sufficiently narrow wave packets.

The constant spinors $u_j$ encode internal degrees of freedom and are normalized as $u_j^\dagger u_j = 1$. One can now introduce the overlap quantities
\begin{equation}
\eta := u_1^\dagger u_2, 
\qquad
\bm{\xi} := u_1^\dagger \bm{\sigma} u_2,
\end{equation}
which control the strength of orbital and spin interference, respectively.

In order to isolate genuine backflow contributions, we impose
\begin{equation}\label{pedy}
p_{1z} > 0, \qquad p_{2z} > 0,
\end{equation}
thereby restricting the spectral support to positive longitudinal momenta. Any local current reversal thus occurs despite strictly positive $p_z$ support, as in standard backflow scenarios.

\subsection{Orbital and spin contributions}

We now separate the orbital and spin parts of the current.
For the orbital contribution one obtains
\begin{equation}
\bm{j}_o = \Im(\phi^\dagger \nabla \phi)
= \sum_{j} |c_j|^2 \bm{p}_j + \bm{j}_{int},
\end{equation}
where the first term describes a uniform drift, while the interference contribution reads
\begin{equation}\label{curqo}
\bm{j}_{int} = (\bm{p}_1 + \bm{p}_2)\, \mathrm{Re}\!\left[c_1^* c_2\, \eta\,
e^{i(\Delta \bm{p}\cdot \bm{r} - \Delta E\, t)}\right],
\end{equation}
with
\begin{equation}\label{deltap}
\Delta \bm{p} = \bm{p}_2 - \bm{p}_1,
\qquad
\Delta E = E_2 - E_1.
\end{equation}

The drift term provides a strictly positive background for $j_z$ and, by itself, does not lead to sign changes.

The spin density obtained from (\ref{super}) and (\ref{phij}) takes the form
\begin{equation}
\bm{S} = \sum_j |c_j|^2 \bm{S}_j + \bm{S}_{int},
\end{equation}
where
\begin{equation}
\bm{S}_j = u_j^\dagger \frac{\bm{\sigma}}{2} u_j,
\end{equation}
and
\begin{equation}
\bm{S}_{int}=\Re \left[c_1^* c_2\, \bm{\xi}\, e^{i(\Delta\bm{p}\cdot\bm{r}-\Delta E\, t)}\right].
\end{equation}

The spin current defined as
\begin{equation}
\bm{j}_s = \nabla \times \bm{S},
\end{equation}
can be decomposed into
\begin{equation}
\bm{j}_s = \nabla \times \Big(\sum_j |c_j|^2 \bm{S}_j\Big)
+ \nabla \times \bm{S}_{int},
\end{equation}
and the first term vanishes since $\bm{S}_j$ are spatially constant. Hence, $\bm{j}_s$ reduces to
\begin{equation}
\bm{j}_s = \nabla \times \bm{S}_{int}.
\end{equation}

This structure of the current has a clear interpretation:
\begin{itemize}
\item the orbital current consists of a drift contribution plus interference corrections,
\item the spin current is generated entirely by interference in the spin sector, and provides a purely oscillatory channel without a classical analogue,
\item both arise from cross terms but act through different mechanisms.
\end{itemize}

\subsection{Local current and oscillatory structure}

Below it is shown that at a fixed spatial point, the $z$-component of the current consists of a constant drift term and a single oscillatory mode. The orbital part reads
\begin{eqnarray}\label{jrt}
j_{o,z}(\bm{r},t)&=&\kappa_o+\alpha_o \cos(\Delta\bm{p}\cdot\bm{r}-\Delta E\, t)\nonumber\\
&&+\beta_o \sin(\Delta\bm{p}\cdot\bm{r}-\Delta E\, t),
\end{eqnarray}
with
\begin{equation}
\kappa_o = |c_1|^2 p_{1z} + |c_2|^2 p_{2z} > 0,
\end{equation}
and
\begin{subequations}\label{albe}
\begin{align}
\alpha_o &= (p_{1z}+p_{2z})\, \Re \big[c_1^* c_2 \eta \big],\\
\beta_o &= -(p_{1z}+p_{2z})\, \Im \big[c_1^* c_2 \eta \big].
\end{align}
\end{subequations}

The spin contribution has an analogous harmonic structure,
\begin{eqnarray}\label{jst}
j_{s,z}(\bm{r},t)&=&\alpha_s \cos(\Delta\bm{p}\cdot\bm{r}-\Delta E\, t)\nonumber\\
&&+ \beta_s \sin(\Delta\bm{p}\cdot\bm{r}-\Delta E\, t),
\end{eqnarray}
with
\begin{subequations}\label{sab}
\begin{align}
\alpha_s &= -\Im\!\left[c_1^* c_2\, (\Delta\bm{p} \times \bm{\xi})_z\right],\\
\beta_s &= -\Re\!\left[c_1^* c_2\, (\Delta\bm{p} \times \bm{\xi})_z\right].
\end{align}
\end{subequations}

Combining both contributions,
\begin{equation}\label{sumsab}
\alpha = \alpha_o + \alpha_s, \qquad \beta = \beta_o + \beta_s,
\end{equation}
the total current can be written as
\begin{eqnarray}\label{jzas}
j_z(\bm{r},t)&=&\kappa_o+\alpha \cos(\Delta\bm{p}\cdot\bm{r}-\Delta E\, t)\nonumber\\
&&+\beta \sin(\Delta\bm{p}\cdot\bm{r}-\Delta E\, t)\\
&=&\kappa_o+\sqrt{\alpha^2+\beta^2}\sin(\Delta\bm{p}\cdot\bm{r}+\theta-\Delta E\, t),\nonumber
\end{eqnarray}
where
\begin{equation}\label{sico}
\sin\theta=\frac{\alpha}{\sqrt{\alpha^2+\beta^2}},\qquad \cos\theta=\frac{\beta}{\sqrt{\alpha^2+\beta^2}}.
\end{equation}
Thus, the current reduces to a single oscillatory mode with effective amplitude $\sqrt{\alpha^2+\beta^2}$.

\subsection{Backflow condition}

The backflow functional becomes
\begin{eqnarray}\label{bafud}
F&=&\int\limits_{t_1}^{t_2}\mathrm{d} t[\kappa_o+\sqrt{\alpha^2+\beta^2}\sin(\Delta\bm{p}\cdot\bm{r}+\theta-\Delta E\, t)]\nonumber\\
&=&\kappa_o(t_2-t_1)+2\frac{\sqrt{\alpha^2+\beta^2}}{\Delta E}\sin\left[\frac{\Delta E}{2}\,(t_2-t_1)\right]\nonumber\\
&&\times\sin\left[\Delta\bm{p}\cdot\bm{r}+\theta-\frac{\Delta E}{2}\, (t_1+t_2)\right],
\end{eqnarray}
where, after integration, we made use of the formula: $\cos x-\cos y=-2\sin\frac{x-y}{2}\,\sin\frac{x+y}{2}$.

To capture the backflow, the integration interval is centered around an optimal time $t_0$ chosen to maximize the negative contribution of the oscillatory term. Setting $t_1 = t_0 - T/2$ and $t_2 = t_0 + T/2$, one obtains
\begin{equation}\label{otim}
t_0 = \frac{1}{\Delta E}\left(\frac{\pi}{2} + \theta + \Delta\bm{p}\cdot\bm{r}\right).
\end{equation}
In general, $\Delta E \neq 0$ whenever $|\bm{p}_1| \neq |\bm{p}_2|$. Energy differences may also arise from transverse momentum components, although they are not the only source of energy splitting.

This choice of $t_0$ aligns the phase of the oscillatory term such that its negative contribution is maximized over the integration interval and gives
\begin{eqnarray}\label{ftr}
F(T)&=&\kappa_o T -2\,\frac{\sqrt{\alpha^2+\beta^2}}{\Delta E}\,\sin\left(\frac{1}{2}\,\Delta E\, T\right)\nonumber\\
&=&\kappa_oT\left[1-\frac{\sqrt{\alpha^2+\beta^2}}{\kappa_o}\cdot\frac{\sin\left(\frac{1}{2}\Delta E\, T\right)}{\frac{1}{2}\Delta E\, T}\right],
\end{eqnarray}
which depends only on the interval length $T$.

Backflow occurs whenever for some $T>0$, one has $F(T)<0$, i.e. when the oscillatory term dominates the drift contribution. For sufficiently small $T$,
\begin{eqnarray}
\frac{\sin\left(\frac{1}{2}\Delta E\, T\right)}{\frac{1}{2}\Delta E\, T}\approx 1,
\end{eqnarray}
so the condition reduces to
\begin{equation}\label{condgen}
\kappa_o<\sqrt{\alpha^2+\beta^2},
\end{equation}
which is also obvious from (\ref{jzas}).

The orbital and spin contributions enter only through the combinations (\ref{sumsab}). In the orbital-dominated regime one has
\begin{equation}
|\alpha_o|,|\beta_o| \gg |\alpha_s|,|\beta_s|,
\end{equation}
while with opposite inequalities spin effects dominate.

A backflow configuration requires $p_{1z},p_{2z}>0$ together with the presence of interference between the modes. In the orbital sector, the emergence of backflow is governed by longitudinal phase differences, controlled by $\Delta p_z = p_{1z}-p_{2z}$, which set the scale of oscillations in the current along the $z$-direction. In the limit $\Delta p_z \to 0$, these longitudinal interference effects become weak, suppressing orbital contributions to current reversal. In contrast, spin-dependent backflow can persist in this regime, as it does not rely on longitudinal phase variation.

While the spin contribution to the Pauli current is well known, its role in generating backflow in minimal two-mode configurations has not been emphasized in this setting. The present construction therefore provides a minimal analytical model in which orbital and spin mechanisms can be separated and analyzed on equal footing.

\section{Explicit examples}\label{explicit}

In this section explicit two-mode configurations will be presented, for which the backflow functional can be evaluated analytically and the condition for negative flux can be verified in a controlled manner.

\subsubsection{Purely orbital backflow}

We begin with the choice $u_1 = u_2$ and
\begin{equation}\label{coeffs}
c_1=\frac{1}{\sqrt{1+\rho^2}}, 
\qquad 
c_2=\frac{\rho\,e^{i\varphi}}{\sqrt{1+\rho^2}},
\end{equation}
where $\rho>0$ is a free parameter. This yields
\begin{equation}\label{ex1_eta}
\eta = 1, \qquad \bm{\xi}=u_1^\dagger \bm{\sigma} u_1.
\end{equation}

Spin orientation can be chosen such that $\bm{\xi}\parallel \Delta\bm{p}$, which eliminates the spin contribution, since from (\ref{sab})
\begin{equation}\label{spin_zero}
\alpha_s=\beta_s=0.
\end{equation}
This corresponds to a state in which the spin quantization axis is aligned with a fixed spatial direction defined by the two-mode momentum difference. Such an alignment can be implemented via standard spin rotations prior to momentum filtering and does not impose additional constraints on the momentum structure of the state. It is introduced here to isolate the orbital contribution within the Pauli current decomposition.

The orbital interference terms remain non-zero and, using (\ref{albe}), take the form
\begin{subequations}\label{akap}
\begin{align}
&\alpha_o = (p_{1z}+p_{2z})\,\Re(c_1^*c_2)
= (p_{1z}+p_{2z})\,\frac{\rho}{1+\rho^2}\cos\varphi,\\
&\beta_o = -(p_{1z}+p_{2z})\,\Im(c_1^*c_2)
= -(p_{1z}+p_{2z})\,\frac{\rho}{1+\rho^2}\sin\varphi,
\end{align}
\end{subequations}
and therefore
\begin{subequations}\label{ex1}
\begin{align}
&\sqrt{\alpha^2+\beta^2}= (p_{1z}+p_{2z})\,\frac{\rho}{1+\rho^2},\\
&\kappa_o = \frac{p_{1z}+\rho^2 p_{2z}}{1+\rho^2},\\
&\tan\theta = -\cot\varphi \;\;\Rightarrow\;\; \theta=\varphi-\frac{\pi}{2}\; (\mathrm{mod}\;\pi).
\end{align}
\end{subequations}

Both $\kappa_o$ and the interference amplitude scale linearly with the longitudinal momenta. As a consequence, achieving backflow requires a suitable tuning of parameters such that the oscillatory term in (\ref{ftr}) becomes comparable to the drift contribution. In this regime, the effect remains relatively weak.

The condition (\ref{condgen}) becomes
\begin{equation}\label{condbea}
\frac{p_{1z}+\rho^2p_{2z}}{\rho(p_{1z}+p_{2z})}<1,
\end{equation}
and can be satisfied by appropriate choices of $\rho$, $p_{1z}$, and $p_{2z}$.

\begin{figure}[h!]
\begin{center}
\includegraphics[width=0.48\textwidth]{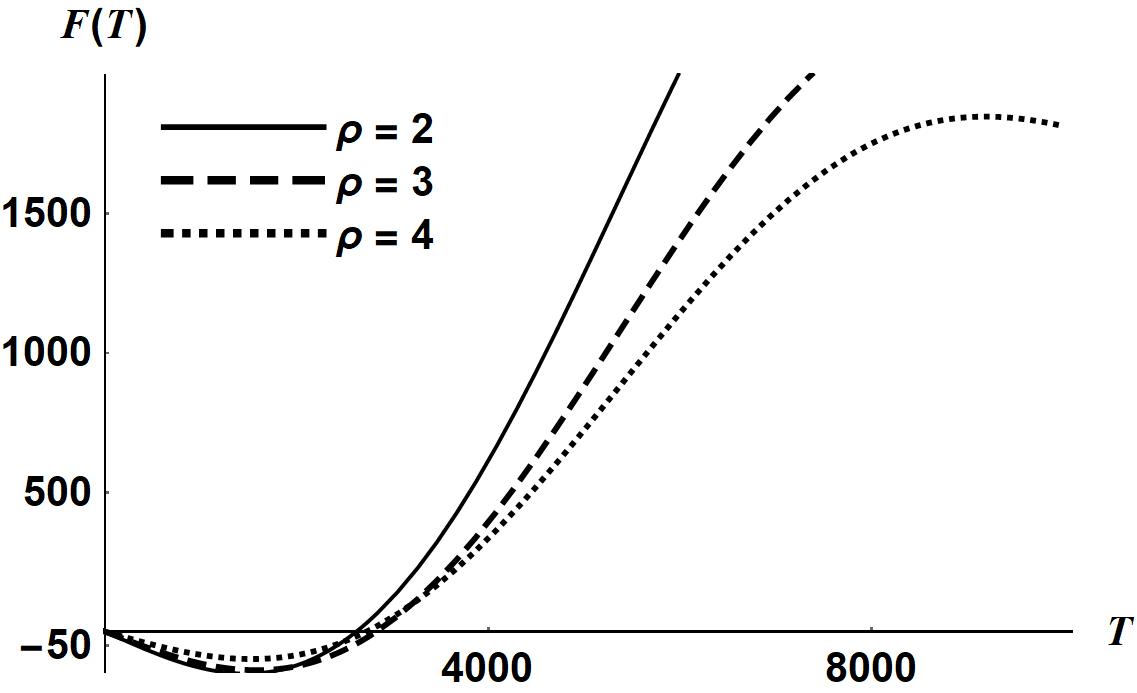}
\caption{Backflow functional as a function of $T$ in the purely orbital case for three chosen values of the superposition parameter $q$. Parameters: $\bm{p}_1=[1,1/10,1]$, $\bm{p}_2=[-1,1,1/9]$, yielding $\Delta E\approx 1.17\cdot 10^{-3}$, $\lambda=-8/9$, and $\rho=3$.}
\label{orbit}
\end{center}
\end{figure}

Setting $p_{2z}-p_{1z}=\lambda p_{1z}$ with constant $\lambda>-1$, one obtains
\begin{equation}\label{condbeb}
\frac{1+ \rho^2(1+\lambda)}{\rho(2+\lambda)}
=\frac{(1-\rho\sqrt{\lambda+1})^2}{\rho(2+\lambda)}+\frac{2\sqrt{\lambda+1}}{2+\lambda}.
\end{equation}

This expression is minimized by choosing $\rho=1/\sqrt{\lambda+1}$, which removes the first term. The remaining contribution can then be made arbitrarily small by suitable parameter tuning.
However, care must be taken to avoid increasing $\Delta E$, since this suppresses the factor $\sin(\tfrac{1}{2}\Delta E T)/(\tfrac{1}{2}\Delta E T)$ in (\ref{ftr}), which controls the temporal decay of interference-induced oscillations.

Figure~\ref{orbit} shows the dependence of the backflow functional on $T$ for representative parameters. The initial regime is dominated by interference-induced negative contributions, but as time increases these oscillatory terms are damped, and the linear drift term associated with the mean momentum becomes dominant. This results in a crossover from transient backflow to asymptotic classical-like transport.

\subsubsection{Pure spin-induced backflow}

In this subsection a configuration is considered in which backflow arises solely from the spin contribution, while orbital interference vanishes. To achieve this the orthogonal spinors are chosen:
\begin{equation}
u_1 =
\begin{pmatrix}
1\\
0
\end{pmatrix},
\qquad
u_2 =
\begin{pmatrix}
0\\
1
\end{pmatrix},
\end{equation}
so that
\begin{equation}
\eta = 0,
\end{equation}
and consequently
\begin{equation}
\alpha_o = \beta_o = 0.
\end{equation}
In this basis the spin interference vector is
\begin{equation}
\bm{\xi} = [1,-i,0].
\end{equation}

We now choose the momenta in the form
\begin{equation}\label{moments}
\bm{p}_1 = [0,0,p],
\qquad
\bm{p}_2 = [-k,0,p],
\qquad p,k>0,
\end{equation}
which differ only by a transverse component in the $x$ direction. This choice preserves the longitudinal drift while introducing a transverse momentum mismatch that couples to the spin structure.

From $\Delta \bm{p} = [-k,0,0]$ it follows that
\begin{equation}\label{iq}
(\Delta\bm{p}\times \bm{\xi})_z = i k,
\end{equation}
and substitution into (\ref{sab}) gives
\begin{align}
\alpha_s &= -\Re\!\left[k\, c_1^* c_2\right],\\
\beta_s &= \Im\!\left[k\, c_1^* c_2\right].
\end{align}
Using (\ref{coeffs}) one obtains
\begin{equation}
c_1^* c_2 = \frac{\rho e^{i\varphi}}{1+\rho^2},
\end{equation}
which leads to
\begin{subequations}\label{abo}
\begin{align}
\alpha_s &=  -k\, \frac{\rho}{1+\rho^2}\, \cos\varphi,\\
\beta_s  &=  k\, \frac{\rho}{1+\rho^2}\, \sin\varphi.
\end{align}
\end{subequations}

Since $\alpha_o=\beta_o=0$, the coefficients in (\ref{jzas}) reduce to
\begin{equation}
\alpha = \alpha_s, \qquad \beta = \beta_s,
\end{equation}
and the longitudinal current becomes
\begin{equation}
j_z(\bm{r},t) = p -k\,\frac{\rho}{1+\rho^2}
\sin(kx - \theta + \tfrac{k^2}{2}\, t),
\end{equation}
with drift term
\begin{equation}
\kappa_o = p.
\end{equation}

In this configuration the orbital part contributes only a constant drift, while all time dependence originates from the spin sector.

The oscillation amplitude is maximal for $\rho=1$, where it equals $k/2$. In this case, the backflow condition is $k>2p$, corresponding to a regime where transverse momentum dominates the longitudinal drift. This shows that spin-induced interference can produce negative flux even when the drift term is fixed. The backflow functional in this case is
\begin{equation}\label{bafus}
F(T)=pT\left[1-\frac{\rho}{1+\rho^2}\cdot\frac{k}{p}\cdot\frac{\sin\left(\frac{k^2T}{4}\right)}{\frac{k^2T}{4}}\right].
\end{equation}

\begin{figure}[h!]
\begin{center}
\includegraphics[width=0.48\textwidth]{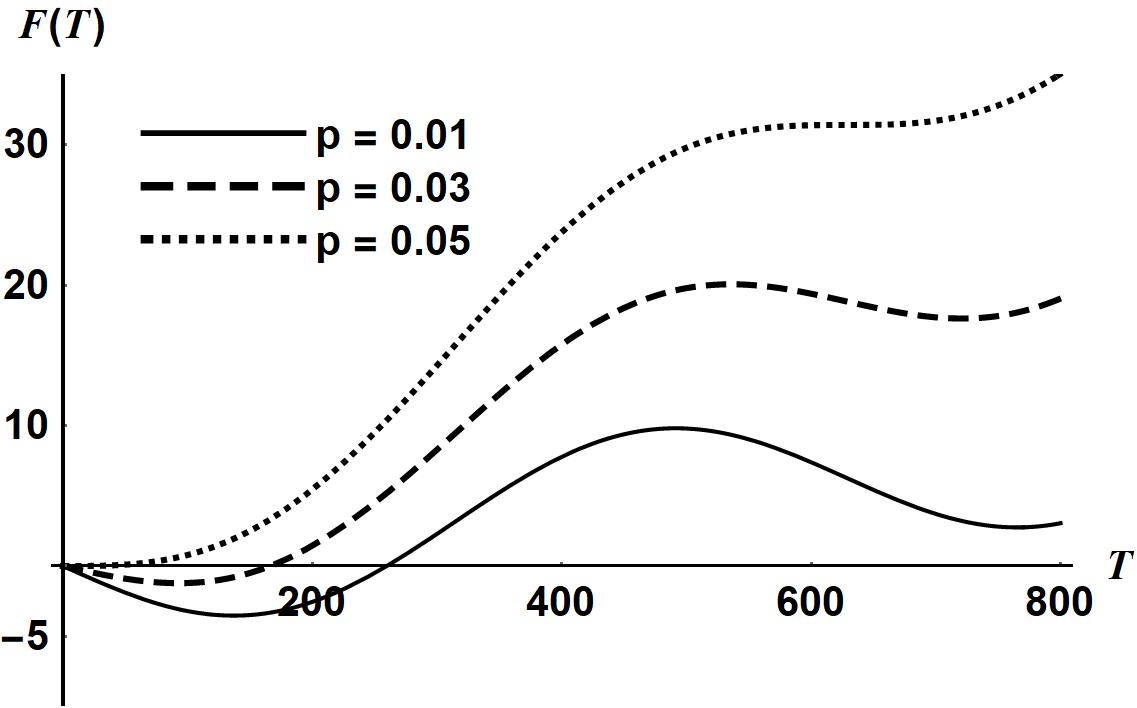}
\caption{Backflow functional as a function of $T$ for three exemplary values of $p$. Negative values arise from the spin contribution, when longitudinal momentum is sufficiently small. Parameters: $k=0.2$, $\rho=1$.}
\label{spin}
\end{center}
\end{figure}

Figure~\ref{spin} shows $F(T)$ for different parameter choices. Negative values arise when the spin contribution exceeds the drift over finite time intervals; otherwise the functional remains positive. Of course the drift term grows with increasing $p$. 

The mechanisms underlying orbital and spin contributions differ in structure. In the orbital case, each plane-wave component contributes a positive drift proportional to $p_{jz}$, and negative flux results from interference-induced suppression of this background. In contrast, the spin contribution is generated entirely by interference through $\bm{j}_s = \nabla \times \bm{S}$,
and vanishes for individual plane waves.

In the present configuration $p_{1z}=p_{2z}$, and the orbital current reduces to a constant background, while all time-dependent contributions to the local flux arise from the spatial structure of the spin density. The spin density becomes spatially non-uniform due to the superposition of spinor components, which generates non-vanishing spin-current contributions through the Pauli term. The resulting current modulation is therefore mediated exclusively by the spin-dependent part of the probability current. Negative flux in this regime is not driven by interference between different longitudinal momenta in the orbital sector, but by the spin contribution to the total current.

The quantity $(\Delta\bm{p}\times \bm{\xi})_z$ determines the strength of this effect and depends only on transverse structure. As a result, spin-induced backflow can persist even when the longitudinal motion of both modes is identical.

Finally, condition (\ref{condgen}) suggests that spin-induced backflow can be enhanced by reducing the drift while maintaining a finite spin amplitude. In the orbital case, both contributions scale with $(p_{1z}+p_{2z})$, which limits independent tuning. 

\section{Controlled enhancement and suppression of backflow}\label{control}

The Pauli current exhibits an interplay between two physically distinct channels: an orbital contribution arising from momentum–phase superposition and a spin contribution originating from the internal structure of the spinor. The relative strength of these channels can be tuned continuously, allowing for both enhancement and suppression of the total backflow. To illustrate this mechanism, we consider the following simple configuration.

If one chooses the momenta
\begin{equation}
\bm{p}_1 = [0,0,p],
\qquad
\bm{p}_2 = [-2p,0,p],
\qquad p>0,
\end{equation}
together with spinors
\begin{equation}
u_1 =
\begin{pmatrix}
1\\
0
\end{pmatrix},
\qquad
u_2 =
\begin{pmatrix}
\cos\delta\\
\sin\delta
\end{pmatrix},
\end{equation}
and equal amplitudes
\begin{equation}
c_1 = c_2=\frac{1}{\sqrt{2}},
\end{equation}
one readily obtains
\begin{equation}
\eta=\cos\delta,\qquad \bm{\xi}=[\sin\delta,-i\sin\delta,\cos\delta],
\end{equation}
and consequently
\begin{equation}\label{sklad}
(\Delta\bm{p}\times\bm{\xi})_z=2 i p \sin\delta.
\end{equation}

This allows to determine the backflow coefficients:
\begin{subequations}\label{becoe}
\begin{align}
\alpha_o &= p\cos\delta,\qquad \beta_o=0, \label{becoeo}\\
\alpha_s &= -p\sin\delta,\qquad \beta_s=0, \label{becoes}\\
\kappa_o &= p. \label{becoek}
\end{align}
\end{subequations}

The resulting current and backflow functional take the form
\begin{subequations}\label{hdr}
\begin{align}
j_z &= p\left[1+(\cos\delta-\sin\delta)\cos(2px+2p^2t)\right],\label{hdra}\\
F(T) &= pT\left[1-(\cos\delta-\sin\delta)\,\frac{\sin(p^2T)}{p^2T}\right].\label{hdrb}
\end{align}
\end{subequations}

\begin{figure}[h!]
\begin{center}
\includegraphics[width=0.48\textwidth,angle=0]{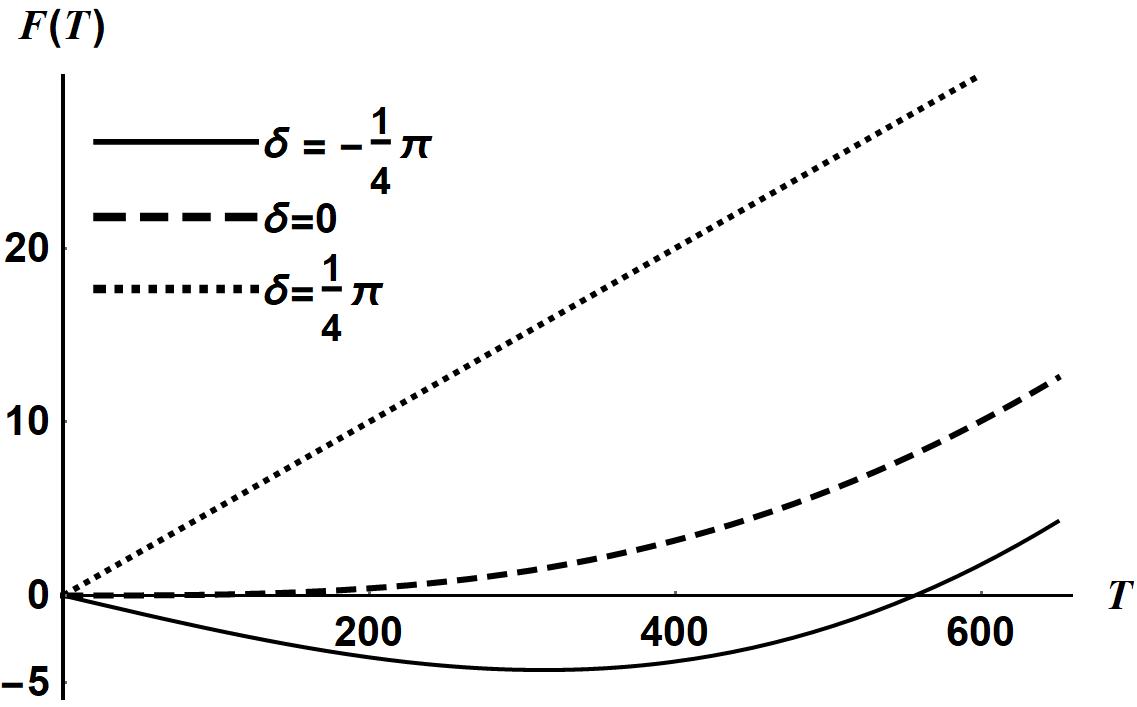}
\caption{Time dependence of the backflow functional $F(T)$ for $p=0.05$ and for three values of the control parameter $\delta$. Varying $\delta$ interpolates between regimes of suppressed (dotted line) and enhanced (solid line) backflow, illustrating explicit control over the interference between orbital and spin contributions.}
\label{inter}
\end{center}
\end{figure}

A particularly transparent situation occurs for $\delta=\tfrac{\pi}{4}$, where the oscillatory contributions cancel in the combined coefficient, leaving only the uniform drift. In contrast, for $\delta=-\tfrac{\pi}{4}$ the oscillatory amplitude is maximized, reaching $\sqrt{2}$ and producing the strongest modulation of the current, provided $p^2T$ remains sufficiently small. Here, “cancellation” refers specifically to the vanishing of the effective oscillatory amplitude in the combined coefficient $\alpha_o+\alpha_s$, rather than a suppression of either channel individually.

Thus, the relative phase between orbital and spin contributions enables a continuous transition between suppression and enhancement of the oscillatory component, allowing controlled tuning of backflow without modifying the momentum distribution. This mechanism has no analogue in scalar quantum mechanics and arises solely from the spinor structure.

The described behaviour is directly reflected in the functional $F(T)$ shown in Fig.~\ref{inter}. In the weak-modulation regime, the functional remains positive and grows monotonically due to the dominant drift term. In contrast, in the strong-modulation regime, pronounced oscillations emerge and can drive $F(T)$ negative over finite time intervals. The latter case corresponds to maximal enhancement of the oscillatory contribution, yielding the most efficient generation of backflow within this setup.

\section{Spatial coarse-graining and robustness of backflow}\label{coarg}

In all considerations so far, the probability current has been evaluated at a fixed spatial point. This choice is natural from a theoretical perspective, as it directly exposes the local interference structure of the wave function. However, any realistic detection scheme has finite spatial resolution, so that pointwise evaluation is replaced by a spatially averaged quantity. It is therefore important to understand how backflow is affected by such coarse-graining.

\subsection{Modified backflow condition}

To formalize this, we replace the local current by a smeared quantity defined as
\begin{equation}\label{coarse}
\widehat{j}_z(\bm{s},t)=\int \mathrm{d}^3 r \, W(\bm{r}-\bm{s})\, j_z(\bm{r},t),
\end{equation}
where $W(\bm{r})$ is a normalized averaging kernel,
\begin{equation}
\int \mathrm{d}^3 r \, W(\bm{r}) = 1,
\end{equation}
encoding the spatial resolution of the measurement. A standard choice is a Gaussian kernel,
\begin{equation}
W(\bm{r}) = \frac{1}{(2\pi\sigma^2)^{3/2}} e^{-\bm{r}^2/(2\sigma^2)},
\end{equation}
where $\sigma$ sets the coarse-graining scale.

This procedure models the finite spatial resolution of realistic detectors. While a Gaussian form is used for explicit calculations, the qualitative behaviour depends only on the presence of a finite-width, sufficiently smooth and rapidly decaying kernel. Such averaging leads to a suppression of high-momentum components. 

Spatial averaging affects drift and interference terms in qualitatively different ways. The drift contribution is either constant or slowly varying on the relevant length scale and is therefore essentially unaffected. By contrast, interference terms contain oscillatory factors of the form
\begin{equation}
e^{i(\Delta \bm{p}\cdot \bm{r} - \Delta E\, t)},
\end{equation}
which are strongly sensitive to spatial integration.

One readily verifies that
\begin{equation}\label{tfw}
\int\mathrm{d}^3r\,e^{i\Delta\bm{p}\cdot\bm{r}}W(\bm{r}-\bm{s})=e^{i\Delta\bm{p}\cdot\bm{s}}\,\widetilde{W}(\Delta\bm{p}),
\end{equation}
where
\begin{equation}\label{werd}
\widetilde{W}(\Delta \bm{p}) = e^{-\frac{1}{2}\sigma^2 |\Delta \bm{p}|^2}.
\end{equation}
The exponential damping factor obtained for the Gaussian case should be understood as a representative example of the general suppression of interference due to finite detector resolution in momentum space.

The coarse-grained current then takes the form
\begin{equation}\label{cgjz}
\widehat{j}_z(\bm{s},t)=\kappa_o+\widetilde{W}(\Delta\bm{p})\sqrt{\alpha^2+\beta^2}\,\sin(\Delta\bm{p}\cdot\bm{s}+\theta-\Delta E\,t),
\end{equation}
and the corresponding coarse-grained backflow functional is defined as
\begin{equation}
\widehat{F} = \int\limits_{t_0-\frac{T}{2}}^{t_0+\frac{T}{2}} \mathrm{d}t\, \widehat{j}_z(\bm{s},t),
\end{equation}
with optimal centering time
\begin{equation}\label{otims}
t_0=\frac{1}{\Delta E}\left(\frac{\pi}{2}+\theta+\Delta\bm{p}\cdot\bm{s}\right).
\end{equation}

Coarse-graining thus introduces the multiplicative factor $\widetilde{W}(\Delta\bm{p})$, so that from (\ref{ftr}) one obtains
\begin{equation}\label{ftrcg}
\widehat{F}(T)
=\kappa_oT\left[1-e^{-\frac{1}{2}\sigma^2 |\Delta \bm{p}|^2}\,\frac{\sqrt{\alpha^2+\beta^2}}{\kappa_o}\cdot\frac{\sin\left(\frac{1}{2}\Delta E\, T\right)}{\frac{1}{2}\Delta E\, T}\right].
\end{equation}

Equivalently, the interference coefficients are renormalized as
\begin{equation}
\alpha \to \alpha\, e^{-\frac{1}{2}\sigma^2 |\Delta \bm{p}|^2},
\qquad
\beta \to \beta\, e^{-\frac{1}{2}\sigma^2 |\Delta \bm{p}|^2},
\end{equation}
while the drift term $\kappa_o$ remains unchanged.

This behavior has a straightforward physical interpretation: spatial averaging suppresses interference fringes whose wavelength falls below the resolution scale $\sigma$, in accordance with general Fourier considerations. This is consistent with the interference-based origin of optical backflow reported in \cite{kot2,he2}. Since the fringe spacing is of order $|\Delta \bm{p}|^{-1}$, suppression becomes significant once
\begin{equation}
\sigma \gtrsim |\Delta \bm{p}|^{-1}.
\end{equation}

Accordingly, the backflow condition (\ref{condgen}) is modified to
\begin{equation}\label{condgenc}
\kappa_o<e^{-\frac{1}{2}\sigma^2 |\Delta \bm{p}|^2}\,\sqrt{\alpha^2+\beta^2},
\end{equation}
which is increasingly difficult to satisfy as $\sigma$ grows.
In the limit of strong coarse-graining,
\begin{equation}
\sigma |\Delta \bm{p}| \gg 1,
\end{equation}
the interference contribution is exponentially suppressed and the current reduces effectively to the drift term,
\begin{equation}
\bar{j}_z \approx \kappa_o > 0,
\end{equation}
so that backflow is eliminated. This behavior is consistent with observations in optical realizations of backflow-like effects, where the phenomenon depends sensitively on the spatial structure of the field \cite{kot1,kot2}.

\subsection{Orbital vs spin contributions under coarse-graining}

Although both orbital and spin contributions acquire the same Fourier suppression factor $\widetilde{W}(\Delta\bm{p}) = e^{-\frac{1}{2}\sigma^2|\Delta\bm{p}|^2}$, their dependence on $\Delta\bm{p}$ is qualitatively different, leading to distinct robustness properties.

For the orbital contribution, the interference amplitude scales as
\begin{equation}
\sqrt{\alpha_o^2+\beta_o^2} \sim (p_{1z}+p_{2z})\,|c_1^*c_2\,\eta|,
\end{equation}
and remains finite as $\Delta\bm{p}\to 0$. However, in this regime the energy splitting $\Delta E$ also becomes small, leading to increasingly slow oscillations. Significant orbital backflow therefore requires a non-negligible momentum separation, in direct competition with the regime where coarse-graining effects are strongest.

In contrast, the spin contribution scales as
\begin{equation}
\sqrt{\alpha_s^2+\beta_s^2} \sim |c_1^*c_2|\,|(\Delta\bm{p}\times\bm{\xi})_z|,
\end{equation}
and is explicitly proportional to $\Delta\bm{p}$. It vanishes for identical modes but can be sustained by purely transverse momentum differences, independently of $\Delta p_z$.
This structure exhibits features reminiscent of azimuthal backflow phenomena in structured optical fields and of the nontrivial spatial structure of the Poynting vector in beam-like electromagnetic fields \cite{zhang,he2,trgaus,ibbc}.

These scalings imply different optimization strategies. In the orbital case, increasing $|\Delta\bm{p}|$ enhances interference but simultaneously strengthens exponential suppression. In the spin case, reducing $|\Delta\bm{p}|$ weakens the intrinsic amplitude itself, since the current is proportional to $\Delta\bm{p}$.

One can conclude that neither contribution is generically more robust under coarse-graining. Instead, robustness is governed by a trade-off between interference strength and spatial resolution effects, which differs between the two channels.

The orbital and spin cases are optimized independently within their respective natural parameter spaces, reflecting the fact that the underlying structures impose different constraints. The comparison presented here therefore concerns the optimal achievable robustness in each case rather than a pointwise comparison at fixed parameter values.

A meaningful comparison must also account for the drift term $\kappa_o$. The relevant quantity controlling backflow is therefore the ratio $\frac{\kappa_o}{A^{\mathrm{eff}}}$ (where $A^{\mathrm{eff}}$ denotes the effective amplitude of the oscillatory part), which determines whether interference can overcome the positive background.

Let us introduce the parametrization
\begin{equation}\label{defpar}
p := p_{1z}, \quad q := \Delta p_z=p_{2z}-p_{1z}, \quad \lambda:= \frac{q}{p},\quad \mu:=\sigma p,
\end{equation}
with $\lambda>-1$.

For the orbital contribution, we consider parallel spinors and analyze the ratio (which is more suitable to plot than its inverse)
\begin{equation}\label{q1}
\frac{\kappa_o}{\sqrt{\alpha_o^2+\beta_o^2}}\,e^{\frac{1}{2}\,\sigma^2|\Delta\bm{p}|^2}.
\end{equation}

Assuming $|\Delta\bm{p}|^2 \simeq |\Delta p_z|^2=q^2$, we define
\begin{eqnarray}\label{fo}
\frac{p_{1z}+\rho^2p_{2z}}{(p_{1z}+p_{2z})\rho}\,e^{\frac{1}{2}\,\sigma^2|\Delta p_z|^2}
&=&\frac{1+\rho^2(1+\lambda)}{(2+\lambda)\rho}\, e^{\frac{1}{2}\,\mu^2\lambda^2}\nonumber\\
&=:&f_o(\rho,\lambda,\mu).
\end{eqnarray}

Backflow is possible when this quantity drops below unity. For fixed $(\lambda,\mu)$, the minimum with respect to $\rho$ is achieved at $\rho=1/\sqrt{1+\lambda}$ (up to an irrelevant phase shift). Substituting this value yields
\begin{equation}\label{foa}
f_o\Big(\frac{1}{\sqrt{1+\lambda}},\lambda,\mu\Big)=\frac{2\sqrt{1+\lambda}}{2+\lambda}\, e^{\frac{1}{2}\,\mu^2\lambda^2}.
\end{equation}

Minimizing over $\lambda>-1$ gives
\begin{equation}\label{mmingam}
\lambda_o=-\frac{3}{2}+\frac{1}{2}\,\sqrt{1+\frac{2}{\mu^2}},
\end{equation}
leading to
\begin{eqnarray}\label{fomin}
f_o^{\mathrm{min}}(\mu)
&=&\frac{4\sqrt{\mu}}{(\sqrt{\mu^2+2}+\mu)^{3/2}}\,e^{\frac{1}{8}(\sqrt{\mu^2+2}-3\mu)^2}.
\end{eqnarray}

A similar optimization can be carried out for the spin contribution. To this goal we consider orthogonal spinors and choose
\begin{equation}
\bm{p}_1 = [0,0,p], 
\qquad 
\bm{p}_2 = [-k,0,p],
\end{equation}
with $p,k>0$ and $\zeta=k/p$. This yields
\begin{eqnarray}\label{q2}
\frac{\kappa_o}{\sqrt{\alpha_s^2+\beta_s^2}}
&&\!\!\!\!\!\!e^{\frac{1}{2}\,\sigma^2|\Delta\bm{p}|^2}
=\frac{(1+\rho^2)p}{\rho k}\, e^{\frac{1}{2}\,\sigma^2k^2}\\
&=&\frac{1+\rho^2}{\rho}\cdot\frac{1}{\zeta}\,e^{\frac{1}{2}\,\mu^2\zeta^2}
=:f_s(\rho,\zeta,\mu).\nonumber
\end{eqnarray}

The minimum is obtained at $\rho=1$ and $\zeta=1/\mu$, i.e. $k=1/\sigma$, giving
\begin{equation}
f^{\mathrm{min}}_s(\mu)=2\sqrt{e}\,\mu.
\end{equation}

\begin{figure}[h!]
\begin{center}
\includegraphics[width=0.48\textwidth,angle=0]{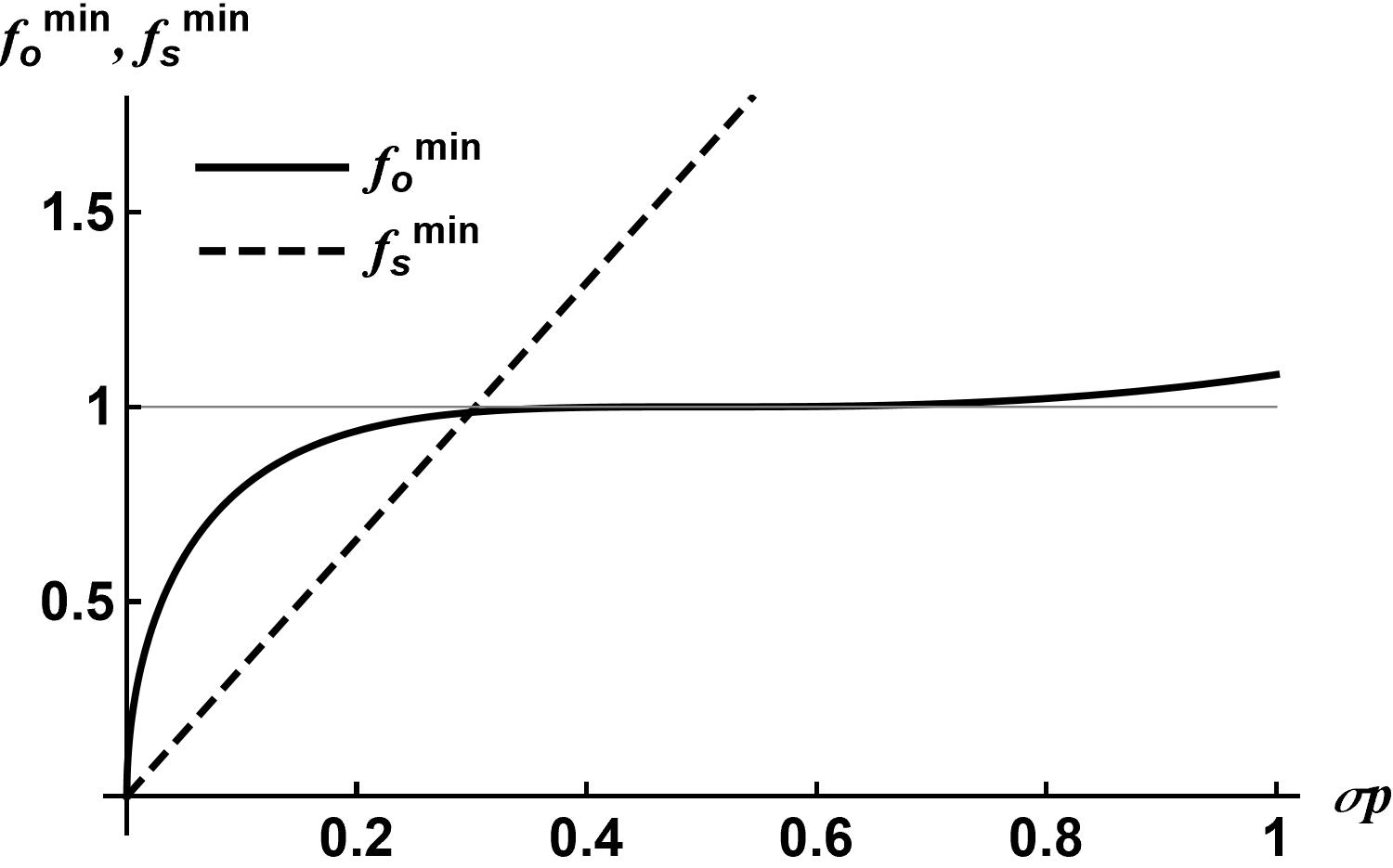}
\caption{Optimal ratios $f^{\mathrm{min}}_o(\mu)$ and $f^{\mathrm{min}}_s(\mu)$ characterizing the survival of orbital and spin-induced backflow under coarse-graining. Values above unity indicate the suppression of backflow. It is evident that the orbital contribution is more sensitive to coarse-graining.}
\label{fmin}
\end{center}
\end{figure}

Thus, spin-induced backflow survives coarse-graining provided
\begin{equation}\label{ineqp}
p<\frac{1}{2\sqrt{e}}\cdot \frac{1}{\sigma}.
\end{equation}

Importantly, the transverse momentum $k$, which controls the spin effect, is independent of the longitudinal scale $p$, so this condition can always be satisfied by reducing $p$ without affecting the mechanism itself.

Figure~\ref{fmin} compares the optimized orbital and spin cases. Both mechanisms are suppressed as the coarse-graining scale increases, but the spin contribution remains more robust over a wider parameter range. 
The curves are obtained under independently optimized parameter choices for each mechanism. This highlights that robustness is governed not by absolute interference strength, but by the interplay between momentum separation and the exponential damping induced by finite spatial resolution.\\

\section{Summary and Conclusions}

The present analysis indicates that quantum backflow in spin-$\tfrac{1}{2}$ systems within the Pauli framework exhibits a richer structure, since the probability current decomposes into orbital and spin-dependent contributions. The spin term introduces an additional interference mechanism with distinct scaling properties relative to the orbital part.

A key result is that negative probability current is not exclusively tied to momentum-space interference. It can also arise from the internal spinor structure, even when the orbital contribution reduces to a uniform drift. Thus, backflow is governed by two independent mechanisms acting on different aspects of the current.

Orbital and spin terms exhibit different scaling with respect to momentum separation and spinor configuration, which leads to distinct optimization strategies. In particular, the spin channel allows interference effects to be controlled without requiring large longitudinal momentum differences, providing a mechanism absent in scalar quantum mechanics.

Spatial averaging associated with finite detector resolution suppresses interference. In the Gaussian coarse-graining model used here, this suppression takes an exponential form, but the qualitative effect is generic and does not depend on the specific choice of averaging kernel. Orbital and spin contributions are affected differently due to their distinct momentum dependence. As a result, the observability of backflow is determined by the interplay between interference strength and measurement resolution, with spin-dominated regimes remaining accessible in parameter ranges where orbital effects are strongly suppressed.

From a broader perspective, spinor degrees of freedom provide an additional control parameter for quantum transport. By tuning spinorial correlations, one can interpolate between regimes of suppressed and enhanced backflow without modifying the momentum distribution.

Several extensions are natural, including multi-mode interference, electromagnetic coupling to the spin sector, and relativistic corrections leading to spin–orbit effects. Identifying experimental implementations in engineered spinor or analogue systems remains an important direction for future work.

The conditions for observable backflow depend on whether criteria analogous to (\ref{condgenc}) are satisfied, with separate implications for orbital- and spin-dominated regimes. This is consistent with recent studies of backflow-like effects in optical settings \cite{elie,daniel,kot2,he2}.

Overall, internal degrees of freedom extend the range of interference-driven transport phenomena beyond the scalar framework.

\end{document}